\documentstyle[12pt]{article}
\newcommand{\ep}{\varepsilon}

\newcommand{\Li}[2]{{\mbox{Li}}_{#1}\left(#2\right)}
\newcommand{\Cl}[2]{{\mbox{Cl}}_{#1}\left(#2\right)}
\newcommand{\Ls}[2]{{\mbox{Ls}}_{#1}\left(#2\right)}
\newcommand{\LS}[3]{{\mbox{Ls}}_{#1}^{(#2)}\left(#3\right)}
\newcommand{\Lsc}[2]{{\mbox{Lsc}}_{#1\!}\left(#2\right)}
\newcommand{\tfrac}[2]{{\textstyle{\frac{#1}{#2}}}}
 
\newcommand{\SN}[3]{S_{#1,#2}\left(#3\right)}

\begin{document}
\renewcommand{\thefootnote}{\fnsymbol{footnote}}

\thispagestyle{empty}

\begin{flushright}
 {MZ-TH/02-05} \\[3mm]
 {DESY~02-031} \\[3mm]
 {hep-th/0203212} \\[3mm]
 {March 2002}
\end{flushright}

 \vspace*{3.0cm}

 \begin{center}
 {\Large \bf
 Geometrical approach to loop calculations \\[2mm]
 and the $\ep$-expansion of Feynman diagrams\footnote{
Contibution to Proceedings of the Conference 
on Computer Particle Physics - CPP2001 (Tokyo, Japan, November 2001).} 
}
 \end{center}
 
\vspace{10mm}

 \begin{center}
 A.~I.~Davydychev$^{a,}$\footnote{On leave from
 Institute for Nuclear Physics, Moscow State University,
 119992, Moscow, Russia. Email address:
 davyd@thep.physik.uni-mainz.de}
 \quad  and \quad
 M.~Yu.~Kalmykov$^{b,}$\footnote{On leave from BLTP, JINR,
                 141980 Dubna, Russia.
Email address: kalmykov@ifh.de}\\

\vspace{10mm}

$^{a}${\em
 Department of Physics,
 University of Mainz, 
 Staudingerweg 7,
 D-55099 Mainz, Germany}
\\

\vspace{3mm}

$^{b}${\em
 DESY--Zeuthen,
Theory Group, Platanenallee 6, D-15738 Zeuthen, Germany}
\end{center}

\vspace{15mm}

\begin{abstract}
\noindent
Some problems related to the structure of 
higher terms of the $\ep$-expansion of Feynman 
diagrams are discussed.
\end{abstract}

\newpage 

\renewcommand{\thefootnote}{\arabic{footnote}}
\setcounter{footnote}{0}

\section{Introduction}

Present accuracy of experimental results \cite{experiment} demands  
calculations of higher-order (two-, three- and, sometimes, four-loop) 
radiative corrections. 
Within dimensional regularization~\cite{dimreg}, with the
space-time dimension $n=4-2\ep$, one needs to 
evaluate higher terms of the $\ep$-expansion of multiloop integrals. 
Dimensional regularization allows one to simultaneously 
regulate the ultraviolet (UV) and infrared (IR) singularities.
The highest power of $1/\ep$ contributing to the UV-singular part 
is connected with the number of loops $L$. 
At the same time, IR-singularities
or the reduction techniques~\cite{ibp,Tarasov}
may introduce additional powers of $1/\ep$.
For this reason, the higher-order terms of the $\ep$-expansion of 
one- and two-loop diagrams become very important 
for multiloop calculations. 
In some cases, one can derive exact analytical results valid 
for an arbitrary space-time dimension $n$, 
usually in terms of various hypergeometric functions~\cite{example,hyper}.
However, even in these cases it may be very difficult to extract 
from these functions
the analytical form of the coefficients of the $\ep$-expansion.
 
In this paper, we review some recent results devoted to construction 
of higher-order coefficients of the 
$\ep$-expansion of (mainly) the single-scale diagrams, whose only mass scale,  
$M$, can be easily factorized as $(M^2)^{-L\ep}$, 
$L$ being the number of loops.
After factorization, the coefficients in $\ep$ for this diagram 
are nothing but a combination of some (in general, irrational) 
numbers. In some cases, these numbers can be numerically calculated for 
each order of $\ep$, 
with a very high accuracy \cite{laporta}. 
Nevertheless, only the exact analytical result 
would allow one to check important properties of quantum field theory, 
like gauge invariance of physical observables and 
the cancellation of the UV-poles~\cite{ATV}.
Among the physically-important cases belonging to this class are 
renormalization group 
(RG) calculations within $\overline{MS}$ 
scheme \cite{RG},
on-shell calculations within QED and QCD \cite{QED} 
and some low-energy electroweak 
processes, where all external (internal) parameters are small with respect 
to internal (external) ones \cite{rho1,rho2}.


\section{Algebraic classification}

Calculation of the RG functions, 
like $\beta$-function or anomalous dimension $\gamma$, 
within $\overline{MS}$ scheme can be reduced to the corresponding 
propagator-type massless diagrams. 
As it was shown in~\cite{ibp}, all three-loop RG functions in 
arbitrary renormalizable models are expressible only in terms 
of $\zeta$-functions. 
The four-loop results for QCD and  five-loop $\beta$-function for
$\phi^4$-model confirm this theorem \cite{RG}. At the higher orders 
(six and more loops) new, non-zeta terms appear \cite{zeta53:1}. 
They can be expressed in terms of the multiple Euler--Zagier
sums \cite{Euler-Zagier},
\begin{equation}
\zeta(s_1,\ldots, s_k; \; \sigma_1,\ldots, \sigma_k)
= \sum_{n_1>n_2> \ldots >n_k>0}\;\;\; \prod_{j=1}^{k}
\frac{(\sigma_j)^{n_j}}{n_j^{s_j}},
\label{euler}
\end{equation}
where $\sigma_j=\pm 1$ and $s_j>0$. 
We will also use a short-hand notation
$$
\zeta_{s_1, \ldots , s_k} \equiv \zeta(s_1, \ldots , s_k; 1, \ldots, 1) \; ,
\qquad U_{a,b} \equiv \zeta(a,b;-1,-1) \; .
$$
For sums of the type~(\ref{euler}), the
{\em weight} ${\bf j}$
can be defined as $\sum_{i=1}^{k}s_i$, whereas
the value of $k$ can be associated with
the {\em depth} (see in \cite{euler-basis,B99}).
For lower cases, these sums correspond to the ordinary $\zeta$-functions.
The six-loop $\beta$-function for the scalar model contains 
$\zeta_{5,3}$ (this was recently confirmed in \cite{zeta53:3}).
At the seven-loop order, a new transcendental number arises~\cite{zeta53:1},
$\zeta_{3,5,3}$.
Note that $\zeta_{5,3}$ and another constant, $\zeta_{7,3}$, appear in the
calculation of anomalous dimensions at ${\cal O}(1/N^3)$ in the
large-$N$ limit \cite{zeta53:2}. 
These results were confirmed and generalized in \cite{sum:even}, where 
it was demonstrated that in odd dimension $3-2\ep$ 
the counterterms contain other constants $U_{a,b}$.
A remarkable property of all these constants is their connection with knots
\cite{zeta53:1,zeta53:2,knot}.
All these results confirm miraculous ``link''
between QFT and the knot theory \cite{knot}:
some Feynman diagrams can be connected with certain knots, so that
the values (Euler--Zagier sums) of Feynman diagrams are also
associated with knots.
A detailed description of the alternating (non-alternating) 
Euler-Zagier sums is presented in Ref.~\cite{euler-basis}. 
All sums of the weight ${\bf j}$ and the depth ${\it k}$, 
satisfying condition $1\le k \le{\bf j}\le7$, are reduced 
to 12 basic elements and their products. 
These irreducible numbers may be taken as~\cite{euler-basis}
\begin{eqnarray}
&& 
\ln 2 \; , \; 
\pi^2 \; , \;
\zeta_3 \; , \;
\zeta_5 \; , \;
\zeta_7 \; , \;
U_{5,1} \; , \;
\zeta(5,1,1;1,1,-1) \; , \;
\zeta(3,3,1;1,1,-1) \; ,   
\nonumber \\ &&  
\left. \left\{
\frac{(-\ln 2)^j}{j!} \left[ 1 - \tfrac{1}{12}j(j-1) 
   \left( \frac{\pi}{\ln 2} \right)^2 \right]
+ \Li{j}{\tfrac{1}{2}} \right\}\right|_{j=4,5,6,7}  .
\end{eqnarray}
In particular, 
$
U_{3,1} =
- 2 \Li{4}{\tfrac{1}{2}} + \tfrac{1}{2} \zeta_4
- \tfrac{1}{12} \ln^4 2 + \tfrac{1}{2} \zeta_2 \ln^2 2 \; .
$
It is interesting that $U_{3,1}$ also arises  
in three-loop massive calculations in QED and QCD 
\cite{QED,rho2} within the on-shell scheme.

Standard Model calculations produce diagrams with other mass distributions. 
Two-loop vacuum diagrams with equal masses \cite{rho1,massive} yield
the transcendental number $\Cl{2}{\frac{\pi}{3}}$, where $\Cl{j}{\theta}$
is the Clausen function~\cite{Lewin},  
$\Cl{2l}{\theta} = \mbox{Im}\; \Li{2l}{e^{{\rm i}\theta}}$
and
$\Cl{2l+1}{\theta} = \mbox{Re}\; \Li{2l+1}{e^{{\rm i}\theta}}$.
The same constant $\Cl{2}{\tfrac{\pi}{3}}$
appears in the one-loop off-shell three-point diagram with 
massless internal lines, in the symmetric case when all external 
momenta squared are equal~\cite{CG}. A generalization
to the $L$-loop ladder case yields $\Cl{2L}{\tfrac{\pi}{3}}$
\cite{UD2}.

At the three-loop level, however, new constants appear \cite{rho2}. 
For their classification, Broadhurst~\cite{B99} has introduced 
the ``sixth root of unity'' basis connected with
\begin{equation}
\zeta\left(\begin{array}{rcr}s_1\;&\ldots&s_k\;\\
\lambda^{p_1}&\ldots&\lambda^{p_k}\end{array}\right) =
\sum_{n_1>n_2> \ldots >n_k>0}\;\;\;
\prod_{j=1}^k \frac{\lambda^{p_j n_j}}{n_j^{s_j}}
\label{sixth}
\end{equation}
where $\lambda=e^{{\rm i}\pi/3}$
and $p_j\in\{0,1,2,3,4,5\}$.
For $p_j \in \{0,3\}$ it coincides with
the Euler--Zagier sums (\ref{euler}).
For the lowest weights ${\bf j}$, the bases consist 
of the following elements:
\begin{eqnarray}
{\bf 1} &:& \pi, \, \ln2, \, \ln3 .
\nonumber \\ 
{\bf 2} &:& \pi^2, \, \ln^22, \, \ln^23, \, 
\Cl{2}{\tfrac{\pi}{3}}, \, \Li{2}{\tfrac{1}{4}}, \, 
\ln2 \ln3, \, \pi \ln2, \, \pi \ln3 .
\nonumber \\ 
{\bf 3} &:&
                \pi\,\Cl{2}{\tfrac{\pi}{3}},\,
                \pi^2\ln2,\,
                \pi^2\ln3,\,
                \Cl{2}{\tfrac{\pi}{3}} \ln2,\,
                \Cl{2}{\tfrac{\pi}{3}} \ln3,\,
                \pi^3,\,
                \Li{2}{\tfrac{1}{4}}\ln2,\,
                \Li{2}{\tfrac{1}{4}}\ln3,\,
                \ln^3 2,\,
\nonumber \\ &&
                \ln^2 2\ln3,\,
                \ln2\ln^2 3,\,
                \ln^3 3,\, 
                \pi\,\Li{2}{\tfrac{1}{4}},\,
                \pi\ln^22,\,
                \pi\ln^23,\,
                \pi\ln2\ln3 , \,
\zeta_3, \,
\Li{3}{\tfrac{i}{\sqrt3}},\,
\Li{3}{\tfrac{\lambda}{2}} .
\nonumber 
\end{eqnarray}
Unfortunately, a large number of elements (more than 4000) makes it
difficult to define the complete basis of ``sixth root of unity'' at 
the weight ${\bf 4}$. So far, only the cases with the depth $k\le 2$
have been examined~\cite{B99}.
One of the remarkable results of Ref.~\cite{B99} is that all
finite parts of three-loop vacuum integrals without subdivergences,
with an arbitrary distribution of massive and massless lines,
can be expressed in terms of four weight-{\bf 4} constants:
$\zeta_4$, $\left[\Cl{2}{\frac{\pi}{3}}\right]^2,U_{3,1}$,
and
$
V_{3,1} = \sum_{p>k>0}
\frac{(-1)^p}{p^3 k}\; \cos\left(\tfrac{2}{3}\pi k\right),
$
which is an essentially new constant.
Note that $\left[\Cl{2}{\frac{\pi}{3}}\right]^2$ also appears
in the two-loop non-planar three-point diagram~\cite{UD3},
when internal lines are massless, whereas all external
momenta squared are off shell and equal.

In Refs.~\cite{sum-log,BB99}, the {\em central binomial sums} were
considered,
$
S(a) \equiv \sum_{n=1}^\infty \frac{(n!)^2}{(2n)!} n^{-a} \; .
$
In particular, in \cite{sum-log} one-fold integral representation 
was established.
It was shown that these sums are connected with the multi-dimensional 
polylogarithm \cite{MDP} of ``sixth root of unity'',
\begin{equation}
\Li{a_1, \ldots, a_k}{\lambda} =
\sum_{n_1>n_2> \ldots >n_k>0}\;\;\;
\frac{\lambda^{n_1}}{n_1^{a_1} \ldots n_k^{a_k}} \; , 
\label{omega}
\end{equation}
and Euler-Zagier sums. 


\section{Geometrical approach}

To predict types of functions (and the values of their arguments)
which may appear in higher
orders of the $\ep$-expansion, a geometrical approach \cite{DD}
happens to be very useful.
Using this approach, the results for {\em all} terms of the
$\ep$-expansion have been obtained for the one-loop two-point
function with arbitrary masses \cite{Crete,D-ep}.
Moreover, {\em all} terms have been also obtained for
the $\ep$-expansion of one-loop
three-point integrals with massless internal lines and
arbitrary (off-shell) external momenta and two-loop
vacuum diagrams with arbitrary masses
\cite{D-ep,bastei_ep}, which are related to each other, due to
the magic connection \cite{DT2}.
All these results have been represented in terms of the
log-sine integrals (see in \cite{Lewin} and below), 
whose angular arguments have a rather
transparent geometrical interpretation (angles of certain
triangles). 
For instance, for the one-loop two-point function with masses
$m_1$ and $m_2$ the relevant angles are (see in~\cite{DD,D-ep})
$\tau'_{01} = \arccos\left(\frac{k^2+m_1^2-m_2^2}{2m_1 \sqrt{k^2}}\right)$ 
and
$\tau'_{02} = \arccos\left(\frac{k^2-m_1^2+m_2^2}{2m_2 \sqrt{k^2}}\right)$,
where $k$ is the external momentum. For single-scale diagram
with $m_1=m_2\equiv m$ and $k^2=m^2$, each of these angles is equal to
$\tfrac{\pi}{3}$.
In more complicated cases, like, e.g., the
three-point function with general values of the momenta
and masses, an arbitrary term of the $\ep$-expansion can
be represented in terms of one-fold angular integrals
whose parameters can be related to the angles accociated
with a three-dimensional simplex \cite{DD,Crete}.

In Ref.~\cite{FKK99}, the on-shell values of
two-loop massive propagator-type integrals have been studied,
and it was observed that the finite (as $\ep\to 0$) parts of all
such integrals without
subdivergences can be expressed in terms of three
weight-{\bf 3} constants,
two for the real part, $\zeta_3$ and $\pi\Cl{2}{\frac{\pi}{3}}$,
and one for the imaginary part, $\pi\zeta_2$.

It is natural to assume that the elements occurring in 
the coefficients of the $\ep$-expansion are connected with some 
properties of diagrams, like the structure of the cut(s). 
In this case, the numbers of elements for higher-order weights
may be essentially reduced. 
Basing on this conjecture,
in Ref.~\cite{FK99} an ansatz was elaborated
for constructing the ``irrationalities'' occurring in the 
$\ep$-expansion of single-scale diagrams involving
cut(s) with two massive particles. 
This construction is closely related to the geometrically-inspired all-order
$\ep$-expansion of the one-loop
propagator-type diagrams \cite{Crete,D-ep}, which was also
used to fix
the normalization factor $\tfrac{1}{\sqrt{3}}$.
The procedure of constructing the ansatz is as follows:
for each given weight ${\bf j}$
the set $\{b_j\}$ of the {\em basic} transcendental numbers
contains (i) all products of the lower-weight elements
$\{b_{j-k} b_k\}$, $k=1,2, \cdots, j-1$ and
(ii) a set of new (non-factorizable) elements
$\{\widetilde{b}_j \}$, which are associated with the quantities
arising in the real and imaginary parts of the polylogarithms
$\Li{j}{e^{{\rm i}\theta}}$ and $\Li{j}{1-e^{{\rm i}\theta}}$,
with $\theta=\tfrac{\pi}{3}$ or $\theta=\tfrac{2\pi}{3}$.
They can be expressed in terms of the Clausen function $\Cl{j}{\theta}$,
log-sine integrals $\Ls{j}{\theta}$ and
generalized log-sine integrals $\LS{j}{k}{\theta}$, 
defined as (see in~\cite{Lewin})
\begin{equation}
\LS{j}{k}{\theta} =   - \int_0^\theta {\rm d}\phi \;
   \phi^k \ln^{j-k-1} \left| 2\sin\tfrac{\phi}{2}\right| \, , \quad
\Ls{j}{\theta} = \LS{j}{0}{\theta} \, .
\label{log-sine}
\end{equation}
Note that $\Ls{2}{\theta}=\Cl{2}{\theta}$.
Therefore, in our case
the non-factorizable part of the basis can be expressed
in terms of the generalized log-sine integrals
of two possible angles,
$\theta=\tfrac{\pi}{3}$ and $\theta=\tfrac{2\pi}{3}$.
To establish the basis of the weight ${\bf 4}$
within such an ansatz, one needs to analyze no more than 100 elements.
It should be noted that this basis is not
uniquely defined, since there are several relations between
polylogarithmic functions $\Cl{j}{\theta}$, $\Ls{j}{\theta}$ and
$\LS{j}{k}{\theta}$ of these arguments.
After excluding all linearly-dependent terms
(see  Appendix~A of \cite{DK2001}), the basis
contains the following non-factorizable constants:
$\Ls{j}{\frac{2\pi}{3}}\bigr|_{j=3,4,5}$,
$\Ls{j}{\frac{\pi}{3}}\bigr|_{j=2,4,5}$,
$\LS{j}{1}{\frac{2\pi}{3}}\bigr|_{j=4,5}$
and $\LS{5}{2}{\frac{2\pi}{3}}$. This set of elements is called
the {\em odd} basis \cite{FK99}. The numerical values of these
constants are given in Appendix~A of \cite{FK99}.
In particular, it was found that the constant $V_{3,1}$ 
can be expressed in terms of the weight-{\bf 4} elements
of the {\em odd} basis,
$
V_{3,1} = \tfrac{1}{3} \left[ \Cl{2}{\tfrac{\pi}{3}} \right]^2
- \tfrac{1}{4} \pi \Ls{3}{\tfrac{2\pi}{3}}
+ \tfrac{13}{24} \zeta_3 \ln 3
- \tfrac{259}{108} \zeta_4
+ \tfrac{3}{8} \LS{4}{1}{\tfrac{2\pi}{3}} \; .
$

By analogy with the
{\em odd} basis introduced in~\cite{FK99}, it is possible to consider
the {\em even} basis connected with the angles $\tfrac{\pi}{2}$
and $\pi$.
Apart from the well-known elements $\pi$, $\ln2$, $\zeta_j$ and the
Catalan's constant
$G=\Cl{2}{\tfrac{\pi}{2}}=\Ls{2}{\tfrac{\pi}{2}}$, this basis 
also contains (up to the weight~{\bf 5})
$\Li{j}{\tfrac{1}{2}}\bigr|_{j=4,5}$ (see also in~\cite{euler-basis}),
$\Ls{j}{\tfrac{\pi}{2}}\bigr|_{j=3,4,5}$,
$\Cl{4}{\tfrac{\pi}{2}}$
and $\LS{5}{2}{\tfrac{\pi}{2}}$.
An example of a physical
calculation where the constant $\Ls{3}{\tfrac{\pi}{2}}$ arises is given in
\cite{ls3_pi/2}.
Instead of $\Li{4}{\tfrac{1}{2}}$ and $\Li{5}{\tfrac{1}{2}}$,
one could take, e.g.,
$\LS{4}{1}{\tfrac{\pi}{2}}$ and $\LS{5}{1}{\tfrac{\pi}{2}}$
(or $\LS{4}{1}{\pi}$ and $\LS{5}{1}{\pi}$) (see Appendix~A of \cite{DK2001}).
The constructed {\em odd} and {\em even} bases
have an interesting property: the number $N_j$ of the basic
irrational constants of a weight ${\bf j}$ satisfies a simple ``empirical'' 
relation $N_j = 2^j$, which has been checked up to weight ${\bf 4}$.
The situation with weight-{\bf 5} bases is discussed below.


\section{Searching for new elements}

A natural generalization of the ansatz proposed in \cite{FK99}
could be the inclusion 
of the generalized (Nielsen) polylogarithms \cite{Nielsen}, 
$S_{a,b}(z)$, where again we consider the cases $z=e^{{\rm i} \theta}$
and $z=1-e^{{\rm i} \theta}$, with $\theta = \frac{\pi}{3}$ or 
$\frac{2\pi}{3}$. It is easy to show that 
$$
\SN{a}{b}{e^{{\rm i} \theta}} = \frac{{\rm i }^{a} (-1)^b}{(a-1)! \,b!}
\int_0^\theta
{\mbox d} \phi \; 
( \theta - \phi )^{a-1}
\left[ \ln \left| 2\sin\tfrac{\phi}{2} \right|
- \tfrac{1}{2}{\rm i} (\pi-\phi)
\right]^b 
+ \SN{a}{b}{1} \; , 
$$
reduces to a combination of the generalized 
log-sine integrals (\ref{log-sine}).
The function $\SN{a}{b}{1-e^{{\rm i} \theta}}$
can be reduced to $\SN{a}{b}{e^{{\rm i} \theta}}$ plus products of
lower-order $\SN{a}{b}{e^{{\rm i} \theta}}$ and powers of 
$\ln\left( 2\sin \tfrac{\theta}{2}\right)$.
Therefore, such Nielsen  polylogarithms  do not generate new elements.
Note that $\SN{a}{b}{1}$ is a combination 
of $\zeta$-functions~\cite{Nielsen}.

Another possibility is the multi-dimensional polylogarithm (\ref{omega}). 
Performing {\sf PSLQ} analysis \cite{PSLQ} with an accuracy of 100 decimals
we established that, up to weight {\bf 5}, all constants (\ref{omega})
are expressible in terms of the {\em odd} basis.  

Let us consider the following hypergeometric function: 
\begin{equation}
_{P+1}F_P\left(\begin{array}{c|}
1+a_1\ep, \ldots, 1+a_{P+1} \ep \\
\tfrac{3}{2} + b \ep,  
2+c_1\ep, \ldots, 2+c_R\ep,
1+d_1\ep, \ldots, 1+d_{P-R-1}\ep
\end{array} ~\frac{z}{4} \right),
\label{PFQ}
\end{equation}
where $0\leq z\leq 4$. 
The one-, two- and three-loop 
Feynman diagrams which are expressible via hypergeometric function 
of this type are given in \cite{example,DK2001}.
The $\ep$-expansion of the function (\ref{PFQ})
can be written as (for details, see Appendix~B of Ref.~\cite{DK2001})
\begin{eqnarray}
&& \hspace*{-7mm} 
\frac{2(1 + 2 b\ep)}{z} 
\left[\prod_{i=1}^{R} (1+c_i\ep) \right]
\sum_{j=1}^\infty \frac{(j!)^2}{(2j)!}  \frac{z^j}{j^{R+1}}  
\Biggl\{ 
1 
- \ep \left[ 
   S_1 T_1
+ \frac{D_1}{j} 
+ 2 b \bar{S}_1
\right] 
+  \ep^2 \Biggl[ 
\frac{1}{2 j^2} \left( D_2 + D_1^2 \right)
\nonumber \\ && 
+  2 b S_1 \bar{S}_1 T_1 
+ \frac{D_1}{j} 
 \left( S_1 T_1
+ 2 b \bar{S}_1  \right)
+ 2 b^2 \left( \bar{S}_2 + \bar{S}_1^2 \right) 
+ \frac{1}{2} S_2  T_2 
+ \frac{1}{2} S_1^2 T_1^2
\Biggr]
+ {\cal O}(\ep^3) 
\Biggr\} \; , 
\end{eqnarray}
where 
$A_k \equiv \sum_{i=1}^{P+1}a_i^k$, 
$C_k \equiv \sum_{i=1}^{R} c_i^k$, 
$D_k \equiv \sum_{i=1}^{P-R-1} d_i^k$ 
and 
$T_k \equiv C_k + D_k - A_k - b^k$. 
It is implied that $C_k=0$ and $\prod_{i=1}^R (1+c_i\ep)=1$
when $R=0$, and $D_k=0$ when $P-R-1=0$.    
Here and below, we also use the short-hand notations
$S_a\equiv S_a(n-1)$ and $\bar{S}_a\equiv S_a(2n-1)$,
where $S_a(n) = \sum_{j=1}^n j^{-a}$ is the harmonic sum. 
In this way, the $\ep$-expansion of the 
hypergeometric function (\ref{PFQ}) 
is reduced to series of the following type:
\begin{equation}
\Sigma_{a_1,\ldots,a_p;
\; b_1,\ldots,b_q;k}^{\; i_1,\ldots,i_p; \;j_1,\ldots,j_q}(z)
\equiv
\sum_{n=1}^\infty \frac{(n!)^2}{(2n)!} \frac{z^n}{n^k}
(S_{a_1})^{i_1}\ldots (S_{a_p})^{i_p}\;
(\bar{S}_{b_1})^{j_1}\ldots (\bar{S}_{b_q})^{j_q}.
\label{binsum}
\end{equation}
For $p=q=0$, the analytical result is available
\cite{KV00}
\begin{equation}
\sum_{n=1}^\infty \frac{(n!)^2}{(2n)!} \frac{z^n}{n^k} 
 = - \sum_{j=0}^{k-2} \frac{(-2)^j}{(k-2-j)!j!} 
( \ln z )^{k-2-j} \LS{j+2}{1}{\theta_z} \;,
\quad
\theta_z\equiv 2\arcsin \frac{\sqrt{z}}{2} \; .
\end{equation}
For the $p+q=1$ terms, the following one-fold integral 
representation can be constructed:
\begin{eqnarray}
\hspace*{-7mm}
&& 
\sum_{n=1}^\infty \frac{(n!)^2}{(2n)!} \frac{z^n}{n^k} 
\left[ \alpha S_1(n-1) + S_1(2n-1)  \right]
=
\frac{2}{(k-2)!} \int_0^{\theta_z}
{\rm d}\theta
\biggl[\ln z - 2 \ln \left(2 \sin \tfrac{\theta}{2} \right) \biggr]^{k-2} 
\nonumber \\ && 
\times
\left\{ 
  \tfrac{1}{2} \Ls{2}{\theta} 
+ \tfrac{1}{2}\theta \ln \left(2 \sin\tfrac{\theta}{2} \right) 
-(1+\alpha) 
\left[\Ls{2}{\pi+\theta} 
+ 
\theta \ln \left(2 \cos\tfrac{\theta}{2} \right) 
\right]
\right\} \; , 
\label{integrals}
\end{eqnarray}
where $k \geq 2$, $0<z<4$ and $\alpha$ is an arbitrary constant.
For all other types of the sums (\ref{binsum}), 
the results are available only for particular values $z=1,2,3$
\cite{DK2001,KV00}.
It should be noted that not all sums (\ref{binsum}) 
can be expressed 
separately in terms of the {\em even} and {\em odd} bases, 
and vice versa, not all basis elements
are expressible in terms of the binomial sums. 

Testing the $\ep$-expansion of a one-loop three-point diagram 
contributing to the $H\to\gamma\gamma$ decay
(see details in \cite{DK2001})
we revealed that new elements $\chi_5$ and $\tilde{\chi}_5$ 
appear 
at the weight-{\bf 5} level in the {\em odd} and {\em even} 
bases, respectively. 
They can be presented as special types of the inverse 
binomial sum (\ref{binsum}),
\begin{eqnarray}
\label{sigma_32}
\chi_5  =   \sum_{n=1}^\infty \frac{(n!)^2}{(2n)!} \; 
\frac{1}{n^2}\; \left[ S_1(n-1) \right]^3
\; ,  \quad 
\tilde{\chi}_5  =   \sum_{n=1}^\infty \frac{(n!)^2}{(2n)!} \; 
\frac{2^n}{n^2}\; \left[ S_1(n-1) \right]^3 \; .
\end{eqnarray}
The r.h.s.\ of Eq.~(\ref{integrals}) gives us an idea about a possible 
integral representation for new functions arising in higher orders 
of the $\ep$-expansion.
Apart from $\LS{j}{i}{\theta}$, we meet 
\begin{eqnarray}
{\mbox{Lsc}}_{i,j} \left(\theta\right) & = & -\int_0^\theta
{\mbox d} \phi \; 
\ln^{i-1} \left| 2\sin\tfrac{\phi}{2} \right|  \; 
\ln^{j-1} \left| 2\cos\tfrac{\phi}{2} \right|  \; .
\label{LSC}
\end{eqnarray}
Some properties of these functions were studied in Appendix~A 
of \cite{DK2001}. 
In particular, a careful {\sf PSLQ} analysis showed that a new element 
for the {\em even} basis should be added, $\Lsc{2,4}{\frac{\pi}{2}}$.
In this way we restore, accidentally, the ``empirical'' relation
$N_j=2^j$ for the ${\bf j}=5$ level of the {\em odd} and {\em even} bases.
However, the class of possible new functions is not restricted only 
by ${\mbox{Lsc}}_{i,j} \left(\theta\right)$ (\ref{LSC}).
We have not tested the harmonic polylogarithms \cite{RV00} 
and their multiple-value generalization \cite{GR2001},
which may be related to the $\ep$-expansion of certain Appell functions 
(see, e.g., in~\cite{bastei_tar}). 
Another possibility is to test all types of the integrals occurring 
in (\ref{integrals}): 
\begin{eqnarray}
&& 
\int_0^\theta {\mbox d} \phi \; 
\phi^k \; 
\ln^i \left| 2\sin\tfrac{\phi}{2} \right|
\ln^j \left| 2\cos\tfrac{\phi}{2} \right|  \; ,
\quad 
\int_0^\theta {\mbox d} \phi \; 
\Ls{2}{\phi} \; 
\ln^i \left| 2\sin\tfrac{\phi}{2} \right|
\ln^j \left| 2\cos\tfrac{\phi}{2} \right|  \; .
\end{eqnarray}
All these issues require a more careful analysis. 
We present here  
one-fold integrals related to our elements 
$\chi_5$ and $\tilde{\chi}_5$
obtained by using a {\sf PSLQ} analysis with an accuracy 
of 200 decimals:
\begin{eqnarray}
&&\hspace*{-7mm}
\int_0^{2\pi/3} {\rm d}\phi \;
\Ls{2}{\phi} \ln^2 \left( 2\sin \tfrac{\phi}{2} \right) 
=
-\tfrac{55}{324} \pi\zeta_2 \Ls{2}{\tfrac{\pi}{3}}
+ \tfrac{1225}{1296} \zeta_2 \zeta_3 + \tfrac{4621}{1296} \zeta_5
+ \tfrac{23}{972} \pi \Ls{4}{\tfrac{\pi}{3}}
\nonumber \\ &&
- \tfrac{5}{18} \pi \Ls{4}{\tfrac{2 \pi}{3}}
- \tfrac{1}{3} \Ls{2}{\tfrac{\pi}{3}} \Ls{3}{\tfrac{2 \pi}{3}}
+ \tfrac{1}{3} \LS{5}{1}{\tfrac{2 \pi}{3}} - \tfrac{1}{48} \chi_5 
\nonumber \\ &&
=  0.744148409838194515377332007\ldots \; ,
\\
&& \hspace*{-7mm}
\int_0^{\pi/2} {\rm d}\phi \;
\Ls{2}{\phi} \ln^2 \left( 2\sin \tfrac{\phi}{2} \right) 
=
- \tfrac{1}{48} \ln^5 2 
+ \tfrac{5}{48}\zeta_2 \ln^3 2 
- \tfrac{35}{128} \zeta_3 \ln^2 2
- \tfrac{5}{8} \Li{4}{\tfrac{1}{2}} \ln 2 
- \tfrac{5}{8} \Li{5}{\tfrac{1}{2}}
\nonumber \\ && 
+ \tfrac{475}{1024} \zeta_2 \zeta_3
+ \tfrac{3379}{2048} \zeta_5\!
- \tfrac{17}{64} \pi \zeta_2 \Ls{2}{\tfrac{\pi}{2}}  
- \tfrac{1}{12} \pi \Ls{4}{\tfrac{\pi}{2}}
+ \tfrac{3}{16} \pi \Cl{4}{\tfrac{\pi}{2}}
- \tfrac{1}{2}  \Ls{2}{\tfrac{\pi}{2}} \Ls{3}{\tfrac{\pi}{2}}\!
- \tfrac{1}{48} \tilde{\chi}_5
\nonumber \\ &&
= 0.656546903118158978863954386\ldots
\end{eqnarray}


\section{Concluding remarks}

Let us return to the idea about the connection between 
the structures occurring in the $\ep$-expansion and 
the structure of the cut(s).
Careful analysis of existing results \cite{set} for the
single-mass-scale diagrams
allows us to formulate the following conjecture \cite{KV00}:
the $\ep$-expansion of diagrams with 
cuts involving 0, 1 or 3 massive lines (or their combinations)
are expressible in terms 
of the {\em even} basis, whereas
the diagrams with a cut involving 2 massive lines (or its
mixing with a 0 or 1 massive cut)  
produce elements of the {\em odd} basis.
For all other cases the problem is open.
Recent results \cite{recent:1} confirm this conjecture. 
We would expect
that the four-loop numerical results in QED \cite{recent:2} can be 
represented in terms of the {\em even} basis.

Another interesting example is
the one-loop triangle diagram $J_3$ with three massive lines
with $m_i=m$,
when all scales (masses and momenta) are equal to each other.
The geometrical approach \cite{DD,Crete} yields the
following representation:
\begin{equation}
J_3(1,1,1;m)\big|_{p_i^2=m^2} = 
- \frac{{\rm i} \pi^{2-\ep}}{m^{2+2\ep}} \Gamma(1+\ep) 2^{1-\ep} 3^{1/2+\ep}
\sum_{j=0}^{\infty} \frac{(-\ep)^j}{(j+1)!}
    \int_0^{\pi/3} \!\!{\rm d}\phi
    \ln^{j+1}\!\left(1\!+\!\frac{1}{8\cos^2\phi}\right)  
\end{equation}
where (see also in Refs.~\cite{vertex:1,vertex:2,vertex:3})
\begin{eqnarray}
\int_0^{\pi/3} {\rm d}\phi
\ln\left( 1 + \frac{1}{8 \cos^2\phi} \right) 
& = & \tfrac{1}{6} \left[ \pi\ln{2} -
\Cl{2}{\tfrac{\pi}{3}} \right] \; ,
\\
\int_0^{\pi/3} {\rm d}\phi
\ln^2 \left( 1 + \frac{1}{8 \cos^2\phi} \right) 
& = & \tfrac{1}{3}  
\Cl{2}{\tfrac{\pi}{3}} \ln 2 
+ \tfrac{1}{3}\pi \ln^2 2 
+ \tfrac{1}{3}\pi \Li{2}{\tfrac{1}{4}}
- \tfrac{1}{2} \Ls{3}{\tfrac{2\pi}{3}}
\nonumber \\ &&
+ 2 {\rm Im}\left[\Li{3}{\tfrac{1}{2} e^{{\rm i}\pi/3}}\right]
- \tfrac{31}{54} \pi \zeta_2 \; .
\end{eqnarray}
One can see that the $\ep$-term is expressed in terms of 
``sixth root of unity'' basis \cite{B99}. However, it is
not sufficient to have only the elements from the 
{\em odd/even} bases. To explain this situation,
we should recall that this vertex diagram also possesses  
an {\em anomalous} threshold. If we put all $p_i^2=p^2$ then
this threshold would correspond to $p^2/m^2 = 3$. 
In other words, this triangle diagram 
has a mixture of an anomalous cut corresponding to $p^2/m^2=3$ 
and two-particle cuts (with respect to each leg)
at $p^2/m^2=4$. A similar situation is also observed  
for the case when $p^2/m^2 = 4$ and $p^2/m^2=9$ cuts 
are present \cite{vertex:3}.

Finally, let us summarize the situation with the {\em even}
and {\em odd} bases.
Although our construction is incomplete, 
it has allowed us to obtain new results for several one-, 
two- and three-loop 
master integrals implemented in different packages \cite{programs}. 
In this way, we found
several interesting relations between generalized log-sine integrals
and multiple zeta functions. Moreover, a new relation between 
${}_3F_2$ and ${}_2F_1$ hypergeometric functions of argument
$\tfrac{1}{4}$ was established \cite{DK2001}.

\vspace{2mm}

{\bf Acknowledgements.} We are grateful to O.V.~Tarasov for useful 
comments.
A.~D.'s research was supported by the Deutsche
Forschungsgemeinschaft.
M.~K.'s research was supported in part by INTAS-CERN grant No.~99-0377.
Partial support from the Australian Research Council grant No.~A00000780
is also acknowledged.
A.~D. is grateful to the Organizers of CPP2001 for their hospitality.

\newpage



\begin{thebibliography}{99}
\bibitem{experiment}
D.~Abbaneo et al.  (ALEPH, DELPHI, L3 and OPAL Collaborations,
LEP Electroweak Working Group, SLD Heavy Flavor and Electroweak Groups),
hep-ex/0112021.

\bibitem{dimreg}
G.~'tHooft and M.~Veltman,
Nucl.\ Phys.\ {\bf B44} (1972) 189;\\
C.G.~Bollini and J.J.~Giambiagi,
Nuovo~Cimento {\bf 12B} (1972) 20; \\
J.F.~Ashmore,  
Lett.\ Nuovo Cim.\ {\bf 4} (1972) 289;\\
G.M.~Cicuta and E.~Montaldi,
Lett.\ Nuovo Cim.\ {\bf 4} (1972) 329.

\bibitem{ibp}
F.V.~Tkachov, Phys.\ Lett.\  {\bf B100} (1981) 65;\\
K.G.~Chetyrkin and F.V.~Tkachov, Nucl.\ Phys.\ {\bf B192} (1981) 159.

\bibitem{Tarasov}
O.V.~Tarasov, Phys.\ Rev.\ {\bf D54} (1996) 6479;
Nucl.\ Phys.\ {\bf B502} (1997) 455.

\bibitem{example}
E.E.~Boos and A.I.~Davydychev,
Theor.\ Math.\ Phys.\ {\bf 89} (1991) 1052;\\
D.J.~Broadhurst, J.~Fleischer and O.V.~Tarasov,
Z.\ Phys.\ {\bf C60} (1993) 287.

\bibitem{hyper}
A.I.~Davydychev,
J.\ Math.\ Phys.\  {\bf 32} (1991) 1052; 
{\bf 33} (1992) 358.

\bibitem{laporta}
S.~Laporta,
Int.\ J.\ Mod.\ Phys.\ {\bf A15} (2000) 5087. 

\bibitem{ATV}
L.V.~Avdeev, O.V.~Tarasov and A.A.~Vladimirov,
Phys.\ Lett.\ {\bf B96} (1980) 94.

\bibitem{RG}
D.I.~Kazakov, O.V.~Tarasov and A.A.~Vladimirov,
Sov.\ Phys.\ JETP {\bf 50} (1979) 521;\\
S.G.~Gorishny et al.,
Phys.\ Lett.\ {\bf B132} (1983) 351; \\
H.~Kleinert et al.,
Phys.\ Lett.\ {\bf B272} (1991) 39; {\bf B319} (1991) 545(E); \\
T.~van Ritbergen, J.A.M.~Vermaseren and S.A.~Larin,
Phys.\ Lett.\ {\bf B400} (1997) 379; \\
K.G.~Chetyrkin,
Phys.\ Lett.\ {\bf B404} (1997) 161; \\
J.A.M.~Vermaseren, S.A.~Larin and T.~van Ritbergen,
Phys.\ Lett.\ {\bf B405} (1997) 327. 

\bibitem{QED}
N.~Gray, D.J.~Broadhurst, W.~Grafe and K.~Schilcher,
Z.\ Phys.\ {\bf C48} (1990) 673; \\
S.~Laporta and E.~Remiddi, Phys.\ Lett.\ {\bf B379} (1996) 283;\\
K.~Melnikov and T.~van~Ritbergen,
Phys.\ Lett.\ {\bf B482} (2000) 99;
Nucl.\ Phys.\ {\bf B591} (2000) 515.

\bibitem{rho1} 
J.~van der Bij and M.~Veltman, Nucl.\ Phys.\ {\bf B231} (1984) 205.

\bibitem{rho2} 
L.~Avdeev et al.,
Phys.Lett. {\bf B336} (1994) 560; {\bf B349} (1995) 597(E); \\
K.G.~Chetyrkin,  J.H.~K\"uhn and  M.~Steinhauser,
Phys.\ Lett.\ {\bf B351} (1995) 331.

\bibitem{zeta53:1}
D.J.~Broadhurst and  D.~Kreimer,
Int.\ J.\ Mod.\ Phys.\ {\bf C6} (1995) 519;
Phys.\ Lett.\ {\bf B393} (1997) 403.

\bibitem{Euler-Zagier}
L.~Euler,
Novi Comm.\ Acad.\ Sci.\ Petropol.\ {\bf 20} (1775) 140;\\
D.~Zagier, 
in {\it Proc. First European
Congress of Mathematics, Paris}, eds. A.~Joseph et al.,
vol.~II, Birkh\"auser, 1994
(Progress in Mathematics, vol.~120), p.~497.

\bibitem{euler-basis}
D.J.~Broadhurst, Open University preprint OUT-4102-62 (hep-th/9604128).

\bibitem{B99}
D.J.~Broadhurst, Eur.\ Phys.\ J.\ {\bf C8} (1999) 311.

\bibitem{zeta53:3}
O.~Schnetz, hep-th/9912149.

\bibitem{zeta53:2}
D.J.~Broadhurst, J.A.~Gracey and  D.~Kreimer,
Z.\ Phys.\ {\bf C75} (1997) 559.

\bibitem{sum:even}
D.J.~Broadhurst and A.V.~Kotikov, Phys.\ Lett.\ {\bf B441} (1998) 345.

\bibitem{knot}
D.~Kreimer, Phys.\ Lett.\ {\bf B354} (1995) 117;
J.\ Knot Th.\ Ram.\ {\bf 6} (1997) 479.

\bibitem{massive}
C.~Ford, I.~Jack and D.R.T.~Jones, Nucl.\ Phys.\ {\bf B387} (1992) 373;
{\bf B504} (1997) 551(E);\\
A.I.~Davydychev and J.B.~Tausk, Nucl.\ Phys.\ {\bf B397} (1993) 123.

\bibitem{Lewin}
L.~Lewin, {\it Polylogarithms and associated functions}
(North-Holland, Amsterdam, 1981).

\bibitem{CG}
W.~Celmaster and R.J.~Gonsalves, Phys.\ Rev.\ {\bf D20} (1979) 1420.

\bibitem{UD2}
N.I.~Ussyukina and A.I.~Davydychev, Phys.\ Lett.\ {\bf B305} (1993) 136.

\bibitem{UD3}
N.I.~Ussyukina and A.I.~Davydychev, Phys.\ Lett.\ {\bf B332} (1994) 159.

\bibitem{sum-log}
Z.~Nan-Yue and K.S.~Williams, Pacific J.\ Math.\ {\bf 168} (1995) 271.

\bibitem{BB99}
D.H.~Bailey and  D.J.~Broadhurst, Math.\ Comp.\ {\bf 70} (2001) 1719 
(math.NA/9905048);\\
J.M.~Borwein, D.J.~Broadhurst and J.~Kamnitzer,
Experimental Math. {\bf 10} (2001) 25 (hep-th/0004153).

\bibitem{MDP}
J.M.~Borwein et al.,
Trans. Amer. Math. Soc. {\bf 353} (2001) 907
(math.CA/9910045).

\bibitem{DD}
A.I.~Davydychev and R.~Delbourgo,
J.\ Math.\ Phys.\ {\bf 39} (1998) 4299.

\bibitem{Crete}
A.I.~Davydychev,
Proc.\ Workshop ``AIHENP-99'', Heraklion, Greece, April 1999 (Parisianou S.A.,
Athens, 2000), p.~219
(hep-th/9908032).

\bibitem{D-ep}
A.I.~Davydychev, Phys.\ Rev.\ {\bf D61} (2000) 087701.

\bibitem{bastei_ep} A.I.~Davydychev and M.Yu.~Kalmykov,
Nucl.\ Phys.\ B (Proc.\ Suppl.) {\bf 89} (2000) 283.

\bibitem{DT2}
A.I.~Davydychev and J.B.~Tausk, Phys.\ Rev.\ {\bf D53} (1996) 7381.

\bibitem{FKK99}
J.~Fleischer, M.Yu.~Kalmykov and  A.V.~Kotikov,
Phys.\ Lett.\ {\bf B462} (1999) 169; {\bf B467} (1999) 310(E).

\bibitem{FK99}
J.~Fleischer and M.~Yu.~Kalmykov, Phys.\ Lett.\ {\bf B470} (1999) 168.

\bibitem{DK2001}
A.I.~Davydychev and  M.Yu.~Kalmykov,
Nucl. Phys. {\bf B605} (2001) 266.

\bibitem{ls3_pi/2}
A.V.~Kotikov and L.N.~Lipatov, Nucl.\ Phys.\ {\bf B582} (2000) 19.

\bibitem{Nielsen}
K.S.~K\"olbig, J.A.~Mignaco and E.~Remiddi, B.I.T. {\bf 10} (1970) 38;\\
R.~Barbieri, J.A.~Mignaco and E.~Remiddi, Nuovo Cim.\ {\bf A11} (1972) 824;\\
A.~Devoto and D.W.~Duke, Riv.\ Nuovo Cim.\  {\bf 7}, No.6 (1984) 1;\\
K.S.~K\"olbig, SIAM J.\ Math.\ Anal.\ {\bf 17} (1986) 1232.

\bibitem{PSLQ}
H.R.P.~Ferguson, D.H.~Bailey and S. Arno,
Math.\ Comp.\ {\bf 68} (1999) 351.

\bibitem{KV00}
M.Yu.~Kalmykov and O.~Veretin, Phys.\ Lett.\ {\bf B483} (2000) 315.

\bibitem{RV00}
E.~Remiddi and J.A.M.~Vermaseren,
Int.\ J.\ Mod.\ Phys.\ {\bf A15} (2000) 725.

\bibitem{GR2001}
A.B.~Goncharov, 
Math.\ Res.\ Lett.\ {\bf 5} (1998) 497; \\
T.~Gehrmann and E.~Remiddi,
Nucl.\ Phys.\ {\bf B601} (2001) 248.

\bibitem{bastei_tar}
O.V.~Tarasov, Nucl.\ Phys.\ B (Proc.\ Suppl.) {\bf 89} (2000) 237;\\
L.G. Cabral-Rosetti and M.A. Sanchis-Lozano,
J.\ Comput.\ Appl.\ Math.\ {\bf 115} (2000) 93.
S.~Moch, P.~Uwer and S.~Weinzierl,  hep-ph/0110083.

\bibitem{set}
D.J.~Broadhurst,
Z.\ Phys.\ {\bf C47} (1990) 115;
{\bf C54} (1992) 599;\\
A.V.~Kotikov,
Phys.\ Lett.\ {\bf B254} (1991) 158;
{\bf B259} (1991) 314;\\
P.N.~Maher, L.~Durand and K.~Riesselmann,
Phys.\ Rev.\ {\bf D48} (1993) 1061; {\bf D52} (1995) 553(E);\\
S.~Bauberger, F.A.~Berends, M.~B\"ohm and M.~Buza,
Nucl.\ Phys.\ {\bf B434} (1995) 383;\\
V.~Borodulin and G.~Jikia, Phys.\ Lett.\ {\bf B391} (1997) 434;\\
S.~Groote, J.G.~K\"orner and A.A.~Pivovarov,
Phys.\ Rev.\ {\bf D60} (1999) 061701; \\
A.~Pelissetto and E.~Vicari, Nucl.\ Phys.\ {\bf B575} (2000) 579.

\bibitem{recent:1}
T.~Gehrmann and E.~Remiddi,
Comput.\ Phys.\ Commun.\  {\bf 141} (2001) 296. 

\bibitem{recent:2}
S.~Laporta,
Phys.\ Lett.\ {\bf B523} (2001) 95.

\bibitem{vertex:1}
U.~Nierste, D.~M\"uller and M.~B\"ohm, Z.\ Phys.\ {\bf C57} (1993) 605.

\bibitem{vertex:2}
A.I.~Davydychev, P.~Osland and L.~Saks,
JHEP {\bf 08} (2001) 050.

\bibitem{vertex:3}
S.~Laporta,
Phys.\ Lett.\ {\bf B504} (2001) 188.

\bibitem{programs}
L.V.~Avdeev, Comput.\ Phys.\ Commun.\ {\bf 98} (1996) 15; \\
J.~Fleischer, M.Yu.~Kalmykov and A.V.~Kotikov,
Proc.\ Workshop ``AIHENP-99'', Heraklion, Greece, April 1999 
(Parisianou S.A., Athens, 2000), p.~231 (hep-ph/9905379); \\
J.~Fleischer and  M.Yu.~Kalmykov,
Comput.\ Phys.\ Commun.\ {\bf 128} (2000) 531; \\
M.~Steinhauser, Comput.\ Phys.\ Commun.\ {\bf 134} (2001) 335.

\end{thebibliography}
\end{document}